\documentstyle[12pt,epsf]{article}
\begin{document}
\title{\bf {Shear Flows and Segregation in the Reaction
$A+B\rightarrow\emptyset$}}
\author{M.J. Howard$^\dagger$ and G.T. Barkema$^\ddagger$}
\date{
{\small\noindent{\it $^\dagger$ Department of Physics, Theoretical
Physics, 1 Keble Road, Oxford, OX1 3NP, U.K. \\ $^\ddagger$ Institute
for Advanced Study, Olden Lane, Princeton, NJ 08540, U.S.A.\\}}
}
\maketitle
\vspace{-10mm}
\begin{abstract}
We study theoretically and numerically the effects of the linear
velocity field ${\bf v}=v_0y{\bf\hat x}$ on the irreversible
reaction $A+B\rightarrow\emptyset$. Assuming homogeneous initial
conditions for the two species, with equal initial densities, we
demonstrate the presence 
of a crossover time $t_c\sim v_0^{-1}$. For $t\ll v_0^{-1}$, the
kinetics are unaffected by the shear and we retain both the effect of
species segregation (for $d<4$) and the density decay rate
$At^{-\alpha}$, where $\alpha=\min({d\over 4},1)$. We calculate the
amplitude $A$ to leading order in a small density expansion for $2\leq
d<4$, and give bounds in $d=4$. However, for $t\gg v_0^{-1}$, the
critical dimension for anomalous kinetics is reduced 
to $d_c=2$, with the density decay rate $Bt^{-1}$ holding for $d\geq
2$. Bounds are calculated for the amplitude $B$ in $d=2$, which
depend on the velocity gradient $v_0$ and the (equal) diffusion
constants $D$. We also briefly consider the case of a non-linear shear
flow, where we give a more general form for the crossover time
$t_c$. Finally,  we perform
numerical simulations for a linear shear flow in $d=2$ with 
results in agreement with theoretical predictions.

\

\end{abstract}
\newpage
\section{Introduction}

Recently there has been considerable activity in the field of
diffusion limited chemical reactions (see [1--9] and references
therein). Many studies have concentrated on the effects
of density fluctuations, especially in the
one and two species reactions $A+A\rightarrow\emptyset$ and
$A+B\rightarrow\emptyset$. In the time dependent case it is well
known that for sufficiently low spatial dimensions these fluctuations
alter the kinetics. For example, in the two species
reaction with equal initial concentrations, 
the densities decay asymptotically as
$t^{-{d\over 4}}$ for $d<4$ \cite {TW,KR,LC} - a slower rate than the
mean field 
$t^{-1}$ result. This is closely related to the phenomenon of
segregation, where the species separate into A and B rich
zones at large times (for $d<4$). These effects are essentially due to
inadequate diffusive mixing of the reactants in low
dimensions. This allows the initial density
fluctuations to persist, leading to
the formation of segregated zones. 

However, exposure of the system to a shear flow will lead to
a modification of these kinetics, as the velocity gradient will allow 
the chemicals to mix more efficiently. 
Hence, we might expect that the critical
dimension for the onset of anomalous kinetics (and the appearance of
segregated zones) might be lowered.
In this paper we attempt to confirm these intuitive expectations by
focusing on a two species reaction in a linear (Couette)
shear flow, in the case where both species have the same diffusion
constant $D$, and the same initial density $n_0$. Previous work on
similar problems includes the study of shear forces
on a binary fluid mixture at criticality (see \cite{KAW} and
references therein). More recently studies have been made
of reaction-diffusion systems where particles with a same species
exclusion rule were subject to a drift \cite{J,J1,IKR}. This exclusion
rule meant that particles of the same species were forbidden from
occupying the same lattice site. New
exponents for the asymptotic density decays were reported, with a
theoretical justification based on the Burgers equation. Note that our
model differs from these cases, both by the nature of the velocity
flow (shear not drift), and by the absence of any exclusion rules.

Although very little work has been done on fluctuations in
reaction-diffusion-shear systems, a considerable amount is known
about diffusion in shear flows
\cite{B,FVDV,MH,BRA}. Exact solutions have been given for
the positional probability distribution of a Brownian particle
released in a linear velocity field. These results will be employed in
the analysis of the following sections. We also mention a paper by Yee
\cite{Y},
where a method was given to obtain an approximate analytic solution to
a general reaction-convection-diffusion equation. However, this
technique is simply a perturbative method for solving the mean
field equations, and as such it takes no account of the microscopic
density fluctuations. So instead we employ a
different method based on a mapping of the microscopic dynamics onto a
quantum field theory \cite{L,LC,me,Doi,Pel}. This allows for a
systematic 
treatment of the density fluctuations using diagrammatic perturbation
theory. In this way we are able to go beyond the traditional rate
equations approach, by using the calculational framework of either an
effective or a full quantum field theory.

In brief we find, for a linear shear flow with velocity gradient
$v_0$, a crossover time of
order $v_0^{-1}$ - at times $t\ll v_0^{-1}$ the behaviour of the
system is essentially unaffected by the shear. However for $t\gg
v_0^{-1}$, the critical dimension for the system is reduced from
$d_c=4$ to $d_c=2$, and 
hence the mean field decay exponent holds in all physically
realisable dimensions. In addition, we note in passing that the 
problem of the single species reaction $A+A\rightarrow\emptyset$
in a shear flow is unlikely to be interesting, as its critical
dimension even without shear is already $d_c=2$ \cite{TW,KR,L}.
 
Finally, we give an outline of the layout of this paper. In section 2
we give simple arguments for the critical dimension 
and crossover times for these systems. The mapping to
a quantum field theory is carried out in section 3, and in section 4
we perform density calculations in the different regimes, using
the field theory formalism. Numerical simulation results are given in
section 5, and we present our conclusions in section 6.

\section{Decay Rates and Crossover Times}

We shall first consider some simple arguments which determine the
density decay rates and crossover times induced by the presence of a
shear flow. If we neglect the role played by fluctuations, then the
mean field result for the densities gives an asymptotic $t^{-1}$
decay. However by using a variant of an argument first put forward by
Toussaint and Wilczek \cite{TW}, we can understand how the density
fluctuations alter this result. At $t=0$, these fluctuations
ensure that
\begin{equation}
|N_a-N_b|_{t=0}\sim({n_0 L^d})^{1\over 2}, \label{fir}
\end{equation}
where $N_a$,$N_b$ are the {\it number} of A,B particles within a
volume $L^d$. Consequently the initial density difference satisfies:
\begin{equation}
|a-b|_{t=0}\sim\left({n_o\over L^d}\right)^{1\over 2}.
\end{equation}
In the absence of a shear flow, after a time $t$, typically only the
{\it 
initial} density excess will remain in a volume of size $(Dt)^{d\over
2}$, as all the other particles in that region will have mutually
annihilated. Hence the number of particles remaining in volume
$(Dt)^{d/2}$ at time $t$ is $(n_0(Dt)^{d\over 2})^{1\over 2}$ (using
equation (\ref{fir})) and we thus 
obtain a $t^{-{d\over 4}}$ density decay \cite{TW,KR,LC}. However, in
our case, the presence of the shear flow means that length
scales parallel to the velocity flow (in the $x$ direction) increase
at a different rate to length scales in perpendicular directions. If
we now specialise to the case of linear shear,
then it is known (from exact solutions for random
walkers in linear velocity flows \cite{B,FVDV,MH}) that this
characteristic length scale grows
as $(Dt[1+{1\over 3}(v_0 t)^2])^{1\over 2}$. Consequently,
after a time $t$, only
the initial density excess will remain in a volume of size
$(Dt)^{d-1\over 2}(Dt)^{1\over 2}[1+{1\over 3}(v_0 t)^2]^{1\over
2}$. As a result we expect the densities to decay asymptotically as
\begin{equation}
a\sim b\sim t^{-{d\over 4}}\left[1+{(v_0 t)^2\over 3}\right]^{-{1\over
4}}. 
\end{equation}
Thus for $t\ll v_0^{-1}$, we retain the $t^{-{d\over 4}}$ decay (for
$d\leq4$), with the mean field $t^{-1}$ exponent applying for
$d>4$. 
However for times very much larger than the crossover time
$t_c\sim v_0^{-1}$, we have a different regime:
\begin{equation}
a\sim b\sim
\cases{t^{-{(d+2)\over 4}}v_0^{-{1\over 2}} & for $d<2$ \cr
t^{-1} & for $d>2$. \cr}
\end{equation}
Note that $d=2$ is the lowest possible dimension for the geometry
of our system, so in practice the mean field decay exponent is always
retained.

We now briefly consider the case of non-linear shear - a
considerably more complicated situation. 
However, we can give some simple arguments which
reveal the
crossover time where we expect the shear to begin to alter the
dynamics. Consider, for example, the non-linear flow
${\bf v}=v_0|y|^n$sgn$(y){\bf\hat x}$ studied in \cite{BRA}. The
shear flow will disrupt a segregated zone around $y=y_0$ when the
typical zone length scale $(Dt_c)^{1\over 2}$ is comparable to the
distance over which the top and bottom of the segregated zone are
sheared apart. The time $t_c$ at which this happens is given by
\begin{equation}
v_0t_c[(Dt_c)^{1\over 2}+y_0]^n-v_0t_cy_0^n\sim (Dt_c)^{1\over 2}.
\end{equation}
We can rewrite this relation for $t_c$ as 
\begin{equation}
v_0t_cy_0^n\left[1+{(Dt_c)^{1\over 2}\over
y_0}\right]^n-v_0t_cy_0^n\sim (Dt_c)^{1\over 2}.
\end{equation}
For the case of a linear shear flow ($n=1$), we have recovered our
earlier result of a crossover time $t_c\sim v_0^{-1}$. However for
$n\not= 1$ and $(Dt_c)^{1\over 2}\ll y_0$, we can expand the bracket
in the above equation. This leads to a crossover time
\begin{equation}
t_c\sim{1\over v_0ny_0^{n-1}}=\left[{d\over
dy_0}(y_0^nv_0)\right]^{-1}, 
\end{equation}
valid when 
\begin{equation}
{D^{1\over 2}\over y_0}\left({1\over v_0ny_0^{n-1}}\right)^{1\over 2}
=\left({D\over v_0ny_0^{n+1}}\right)^{1\over 2}\ll 1.
\end{equation}

\section{The Field Theory Approach}

In order to perform more quantitative calculations for the densities,
we need a systematic way of including the effect of microscopic
density fluctuations. One way in which this can be achieved is by
mapping the microscopic dynamics onto a quantum field theory.
The first step in this process is
to write down a master equation for the microscopic dynamics. This can
then be recast in a second quantised formalism, which may in
turn be mapped onto a path integral. All of these steps have been
described in detail elsewhere \cite {L,LC,me,Doi,Pel}, so we simply
give the 
resulting field theoretic action. Defining the fields $\phi$ and
$\psi$ in terms of the continuous c-number fields $a$ and $b$, where
$\phi={1\over 2}(a+b)$ and $\psi={1\over 2}(a-b)$, we have: 
\begin{eqnarray}
& & S=\int dxdyd^{d-2}zdt\left[2\bar\psi(\partial_t+v(y)\partial_x-
D\nabla^2)\psi+2\bar\phi(\partial_t+v(y)\partial_x-D\nabla^2)\phi
\right. \nonumber \\ & & \qquad\qquad\qquad\qquad\quad\left.
+2\lambda\bar\phi(\phi^2-\psi^2)+\lambda(\bar\phi^2-\bar\psi^2) 
(\phi^2-\psi^2)-2n_0\bar\phi\delta(t)\right] \label{action}
\end{eqnarray}
where $\bar\psi$ and $\bar\phi$ are the response fields. The vertices
for this field 
theory are shown in figure 1, where the propagators for the $\phi$
field are solid lines, and the $\psi$ propagators are dotted lines.
Notice that
if we neglect the quartic terms in the action, and integrate out
the response fields, then we recover the classical (mean field)
equations:
\begin{eqnarray}
& & {\partial\psi\over\partial t}+v(y){\partial\psi\over\partial
x}=D\nabla^2\psi \\
& & {\partial\phi\over\partial t}+v(y){\partial\phi\over\partial
x}=D\nabla^2\phi-\lambda(\phi^2-\psi^2)+n_0\delta (t).
\end{eqnarray}

To derive the form of the Green functions we again specialise
to a linear velocity field, i.e. $v(y)=v_0 y$. The equation for the
Green functions $G=G_{\psi\bar\psi}=G_{\phi\bar\phi}$ is:
\begin{equation}
2\left({\partial\over\partial t}+v_0y{\partial\over\partial
x}-D\nabla^2\right) G = \delta(x-x')\delta(y-y')\delta({\bf z}
-{\bf z'})\delta(t-t'). \label{Greq}
\end{equation}
This equation has been solved elsewhere \cite{B,FVDV,MH,BRA},
in the context of the position probability distribution for a random
walker released from the origin in a linear shear flow:
\begin{eqnarray}
& & G=(4\pi D(t-t'))^{-{d\over 2}}\left({3\over
(v_0(t-t'))^2+12}\right)^{1\over 2}\times \label{Green} \\
& & \qquad\quad\exp\left(-{3(x-x'-{1\over
2}v_0(y+y')(t-t'))^2\over 
D(t-t')[(v_0(t-t'))^2+12]}-{(y-y')^2\over 4D(t-t')}-{({\bf z}-{\bf
z'})^2\over 4D(t-t')}\right). \nonumber
\end{eqnarray}
Note that for $v_0(t-t')\ll 1$ the Green function is the essentially
same as for a shearless system (with $v_0=0$). Physically
this result corresponds to the dominance of diffusion over shear on
short enough time scales.

Finally, we can use the field theory to define two classical (tree
level) quantities. The first of these is the classical density
$\langle\phi\rangle_{cl}$, which is the sum of tree diagrams
contributing to $\langle\phi\rangle$ (see figure 2, where
$\langle\phi\rangle_{cl}$ is represented by a wavy solid line). This
sum can easily be evaluated \cite{L,LC}, giving
\begin{equation}
\langle\phi\rangle_{cl}={n_0\over 1+n_0\lambda t},
\end{equation}
the same result as would be found by solving the mean field rate
equations. The second classical quantity is the response function,
which is defined to be the sum of all possible tree diagrams connected
to a single propagator. The nature of the vertices ensures that the
$\psi$ response function is just the $\psi$ propagator, whereas the
$\phi$ response function is given by the diagrammatic sum shown in
figure 3 (where it is represented by a thick solid line).

\section{Field Theory Density Calculations}

Using the formalism presented in section 3, we are now in a position
to 
calculate the densities in the two distinct time regimes. We consider
first the case of $2<d\leq 4$ and $v_0t\ll 1$, where an effective
field theory can be developed. However such a theory turns out to be
almost identical to that previously used by Lee and Cardy
to study the same two species reaction, but without
shear (see \cite{LC} for details). The only fundamental difference
lies in the slightly modified form of the
Green functions (\ref{Green}).
However, they are still sufficiently similar to
those of Lee and Cardy that our results, to leading order in a small
density expansion, reproduce those of \cite{LC}, and are independent
of $v_0$. Hence, quoting from \cite{LC}, we have, for $2<d<4$ and
$(n_0\lambda)^{-1}\ll t\ll v_0^{-1}$:
\begin{equation}
\langle a\rangle,\langle b\rangle\sim{\Delta^{1\over 2}\over
(2\pi)^{1\over 2}(8\pi)^{d\over 4}}(Dt)^{-{d\over 4}}, \label{res}
\end{equation}
where $\Delta$ is found by summing the diagrams shown in figure 4,
giving $\Delta=2n_0$ to leading order in a small density expansion.
Note that we require $t\gg (n_0\lambda)^{-1}$ so that the
coarse-graining required for the calculation of the initial term
$\Delta$ is valid. Strictly for this and subsequent results, we
must also have $Dt\gg n_0^{-2/d}$ - i.e. the particles must have had
time to ``find'' each other by diffusion. For $d=4$ we have the upper
bound 
\begin{equation}
\langle a\rangle_{upper},\langle b\rangle_{upper}=\left({1\over
2\lambda_{eff}}+\sqrt{{1\over(2\lambda_{eff})^2}+{\Delta\over 
4(8\pi D)^2}}\right)t^{-1},
\end{equation}
where $\lambda_{eff}$ is an effective coupling constant found by
summing the diagrams in figure 5, giving for $v_0t\ll 1$:
\begin{equation}
\lambda_{eff}=\lambda\left(1+{2\Lambda^{d-2}\lambda\over
(8\pi)^{d/2}(d-2)D}\right)^{-1}.
\end{equation}
Here $\Lambda$ is a cutoff which is needed to keep the loop
integrals finite.
The lower bound for the densities in $d=4$ is given by
\begin{equation}
\langle a\rangle_{lower},\langle b\rangle_{lower}={\Delta^{1\over
2}\over 8\sqrt{2}\pi^{3\over 2}}(Dt)^{-1}.
\end{equation}
Finally, it is important to note that these calculations are valid
only in the regime $(n_0\lambda)^{-1}\ll t\ll v_0^{-1}$ and $Dt\gg
n_0^{-2/d}$. Hence, for
small enough $n_0$ and large enough $v_0$, these conditions will not
be satisfied and the above results will not be applicable.

\subsection{$v_0t\ll 1$ and $d=2$}

Power counting on the full field theoretic action (\ref{action})
gives $d=2$ as a critical dimension for the system (when $v_0t\ll 1$).
In this case (of the lowest physically possible dimension), we
must consider the full theory, as given by the action
(\ref{action}). The renormalisation is similar to that previously
developed in \cite{L,LC,me}, where more details may be found. In
particular the field theory remains simple in that
diagrams cannot be drawn which dress the propagators. Consequently the
bare propagators are the full propagators and both $v_0$ and $D$
are not renormalised. Furthermore the simpler form of $G$ for
$v_0t\ll 1$ again ensures that the results for the
$\langle\phi^2\rangle$ and $\langle\psi^2\rangle$
loops are unchanged from the shearless case \cite{LC}. Thus the
primitively divergent vertex function $\bar\lambda (k,t)$ (figure 6),
and resulting $\beta$ function, remain unaltered. Note that we have
now rescaled the couplings to absorb the diffusion constant $D$:
$\bar\lambda=\lambda/D$.

The Callan-Symanzik equation for the densities is slightly modified to
read 
\begin{equation}
\left[2\bar t{\partial\over\partial\bar t}-2\bar
v_0{\partial\over\partial 
\bar v_0}-dn_0{\partial\over\partial
n_0}-d\Delta{\partial\over\partial\Delta}+\beta(g_R)
{\partial\over\partial
g_R}+d\right]n(\bar t,g_R,n_0,\bar v_0,\Delta)=0,
\end{equation}
where $\bar v_0=v_0/D$, $\bar t=Dt$, and $g_R$ is the
dimensionless renormalised coupling. The
solution can be found by the method of characteristics 
\begin{equation}
n(\bar t,g_R,n_0,\bar v_0,\Delta)=(\kappa^2\bar t)^{-{d\over
2}}n(\kappa^{-2},\tilde g_R,\tilde n_0,\tilde{\bar v_0},\tilde\Delta),
\end{equation}
where $\tilde n_0=n_0(\kappa^2\bar t)^{d\over 2}$,
$\tilde\Delta=\Delta(\kappa^2\bar t)^{d\over 2}$, and $\tilde{\bar
v_0}=\bar v_0(\kappa^2\bar t)$. Furthermore, at large enough times
(but still such that $v_0 t\ll 1$), the 
running coupling $\tilde g_R$ goes to zero as $(\ln\bar t)^{-1}$
for $d=2$ \cite{L}. The leading order result is given by (\ref{res}),
and following 
\cite{LC}, we make the assumptions that higher order terms are
both independent of $n_0$, $\bar v_0$ (and thus of $\tilde n_0$,
$\tilde{\bar v_0}$), and that they
diverge no more quickly than $\tilde\Delta^{1\over 2}$, for large
$\tilde\Delta$. Consequently, if these assumptions
are valid, then the densities in $d=2$ will be given by expression
(\ref{res}), with corrections which are suppressed by at least a
factor of $(\ln\bar t)^{-1}$.

\subsection{$v_0t\gg 1$}

At truly asymptotic times $t\gg v_0^{-1}$, we can use the Green
functions $G=G_{\psi\bar\psi}=G_{\phi\bar\phi}$ (\ref{Green}) to
motivate a new assignment of dimensions for the
parameters appearing in the action (\ref{action}). Firstly we can see
{}from (\ref{Green}) that $[x]\sim[Dv_0^2t^3]^{1\over 2}$. If we
ignore the $\partial^2_x$ terms in the action 
(corresponding to the neglect of diffusive motion in the $x$
direction) and assign $[\bar v_0]\sim k^0$, then we can give the
following naive dimensions:
\begin{equation}
[x]\sim k^{-3}\quad[y],[z]\sim k^{-1}\quad[\bar t]\sim k^{-2}\quad
[\bar\lambda]\sim k^{-d}\quad [\bar\phi],[\bar\psi]\sim k^0\quad
[\phi],[\psi]\sim k^{d+2}.
\end{equation}
This suggests that a full field theory analysis
using the action (\ref{action}), with subsequent renormalisation,
becomes necessary only at dimension $d=0$.

Consequently, following \cite{LC}, we must now construct a new
effective field theory, valid for $d>0$. The first step in this
process is to determine which initial parameters are relevant.
For an initial term of the 
type $(\Delta^{(m,n)}/m!n!)\bar\phi^m \bar\psi^n|_{t=0}$, we must
therefore consider the dimensions of the coupling
$[\bar\lambda^{m+n}\Delta^{(m,n)}]\sim
k^{-d(m+n)+d+2}$. The power of $\bar\lambda$ follows from considering
the number of vertices needed to attach the initial term to a given
diagram. These terms will be relevant when
$d<2/(m+n-1)$. Hence, if we have $m+n=1$, then such an initial term is
relevant for all $d$. The case of $m=1$ corresponds to the initial
density, whereas the generation of an $n=1$ term is forbidden by the
invariance of the system under a transformation exchanging
$A\leftrightarrow B$,
i.e. $(\phi,\bar\phi,\psi,\bar\psi)\rightarrow
(\phi,\bar\phi,-\psi,-\bar\psi)$. The only other important
initial terms are those with $m+n=2$, which are marginal in $d=2$.
In fact we need only consider an extra initial term of the type
$(\Delta^{(0,2)}/2)\bar\psi^2$, as $\Delta^{(1,1)}$ is forbidden by
symmetry, 
and $\Delta^{(2,0)}$ is suppressed (it can only act as a source of
noise through a $\phi$ response function, which we assume to be
heavily damped, as in
\cite{LC}). So, for $d\leq2$, we are led to the construction of an
effective field theory with an extra initial term, whereas for $d>2$
all initial terms (except the initial density) are irrelevant and
hence the rate equation
approach can be employed. In what follows we shall develop the
effective field theory only in $d=2$, as the system cannot be realised
in a lower dimension.

Turning to the calculation of
$\Delta^{(0,2)}=\Delta$, we again need
to sum the set of diagrams shown in figure 4, which for $t\gg
(n_0\lambda)^{-1}$ gives $\Delta=2n_0$ to lowest order in a
small density expansion \cite{LC}.
Aside from this term our action is now linear
in the response fields, which we integrate out to yield the
equations of motion:
\begin{equation}
{\partial\phi\over\partial t}+
v_0y{\partial\phi\over\partial 
x}=D\nabla^2\phi-\tilde\lambda_{eff}(\phi^2-\psi^2) \label{aem}
\end{equation}
\begin{equation}
{\partial\psi\over\partial t}+
v_0y{\partial\psi\over\partial x}=D\nabla^2\psi,
\end{equation}
where $\tilde\lambda_{eff}$ is a new effective reaction rate constant,
found by summing the diagrams shown in figure 5, giving
\begin{equation}
\tilde\lambda_{eff}=\lambda\left(1+{4\sqrt{3}\Lambda^d\lambda\over
(8\pi)^{d/2}d\bar v_0 D}\right)^{-1},
\end{equation}
in the limit $v_0t\gg 1$. Note that this result ensures that the
density 
amplitudes are $v_0$ dependent even above the upper critical
dimension $d_c=2$.
If we now average equation (\ref{aem}) over the initial
conditions, then we have
\begin{equation}
{d\over
dt}\langle\phi\rangle=-\tilde\lambda_{eff}\langle\phi^2\rangle+\tilde
\lambda_{eff}\langle\psi^2\rangle, \label{avf}
\end{equation}
since $\nabla^2\langle\phi\rangle=0$ and $\partial_x 
\langle\phi\rangle=0$. However we can see from a diagrammatic
expansion for $\langle\phi\rangle$ (figure 7) that the only diagram
contributing to the value of $\langle\psi^2\rangle$ in (\ref{avf}) is
the single $\psi$ loop, which we now evaluate:
\begin{equation}
\langle\psi^2\rangle=\Delta\int dxdyd^{d-2}z\,G(x,y,{\bf z},t)^2
\qquad\qquad\qquad\qquad\qquad\qquad\qquad\qquad\qquad \
\end{equation}
\begin{equation}
\qquad =\Delta\int {dxdyd^{d-2}z\over (4\pi 
Dt)^{d}}\left({3\over (v_0t)^2+12}\right)
\exp\left\{-{6(x-{1\over 2}v_0yt)^2\over Dt[(v_0t)^2+12]}-{y^2\over
2Dt}-{{\bf z}^2\over 2Dt}\right\}. \nonumber
\end{equation}
For $v_0t\gg 1$ and $d=2$, this gives the result:
\begin{equation}
\langle\psi^2\rangle={\sqrt{3}\Delta\over 16\pi Dv_0 t^2}.
\end{equation}
We can now find an upper bound solution to (\ref{avf}) by replacing
$\langle\phi^2\rangle$ by $\langle\phi\rangle^2$ (see \cite{LC} for a
proof). Calling
this upper bound solution $f$, we have for $v_0t\gg 1$ 
\begin{equation}
{df\over
dt}=-\tilde\lambda_{eff}f^2+{\tilde\lambda_{eff}\Delta\sqrt{3}\over 
16\pi Dv_0 t^2}.
\end{equation}
This can be solved by making the substitution $f=\dot u/\lambda u$,
giving the upper bound
\begin{equation}
\langle\phi\rangle\leq f=\left[{1\over
2\tilde\lambda_{eff}}+\sqrt{{1\over(2\tilde\lambda_{eff})^2} 
+{\Delta\sqrt{3}\over 16\pi Dv_0}}\right]t^{-1}, \label{limu}
\end{equation}
in $d=2$.
However, we can also find a lower bound for $\langle\phi\rangle$ by
noting that $\phi({\bf x},t)\geq|\psi({\bf x},t)|$, or equivalently
that $a({\bf x},t)$ and $b({\bf x},t)$ are everywhere non-negative. We
can prove this rigorously using the effective field theory equations
for $a$ and $b$:
\begin{equation}
{\partial a\over\partial t}=D\nabla^2 a-v_0y{\partial a\over\partial
x}-\tilde\lambda_{eff}ab 
\end{equation}
\begin{equation}
{\partial b\over\partial t}=D\nabla^2 b-v_0y{\partial b\over\partial
x}-\tilde\lambda_{eff}ab.
\end{equation}
We now assume that the (smooth) fields $a$ and $b$ are initially
everywhere non-negative. Suppose, at a later time,
$a=0$ at a point, then $a>0$ locally around the point, implying that
it is a local minimum. Hence $\nabla^2 a>0$ and $\partial_x a=0$,
meaning that $\partial_t a>0$. For a region of $a=0$,
then we have $\partial_t a=0$ inside the region. On its boundaries we
have $\nabla^2 a>0$ and $\partial_x a=0$, giving
$\partial_t a>0$ at these points. As a result the fields cannot pass
through zero and will remain non-negative.

Since we have $\phi\geq|\psi|$, it follows that
$\langle\phi\rangle\geq\langle|\psi|\rangle$. In addition, at long
enough times, we expect $\psi$ to have a
normal distribution - a result of $\psi$ satisfying a
(modified) diffusion equation. Consequently (as in \cite{LC}), we have
\begin{equation}
P[\psi(t)]\propto\exp\left\{-{\psi(t)^2\over
2\langle\psi(t)^2\rangle}\right\}, \label{norm}
\end{equation}
and therefore
\begin{equation}
\langle\phi\rangle\geq\langle|\psi(t)|\rangle=\sqrt{{2\over\pi}
\langle\psi(t)^2\rangle}=\sqrt{ 
\Delta D\sqrt{3}\over 8\pi^2v_0}(Dt)^{-1},
\label{limb}
\end{equation}
for $d=2$. However since $\langle g^2\rangle\geq\langle g\rangle^2$
for any real $g$, then we also have
\begin{equation}
\langle\phi-|\psi|\rangle^2\leq\langle(\phi-|\psi|)^2\rangle=
\langle\phi^2\rangle+\langle\psi^2\rangle -2\langle\phi|\psi|\rangle,
\end{equation}
and using $\phi\geq|\psi|$ this gives us (in $d=2$):
\begin{equation}
\langle\phi-|\psi|\rangle^2\leq\langle\phi^2\rangle
-\langle\psi^2\rangle= -{1\over
\tilde\lambda_{eff}}\langle\dot\phi\rangle\sim O(t^{-2}).
\end{equation}
In other words, for $d=2$, the bound on the 
corrections is of the same order as the density. Consequently we
cannot say that the density asymptotically approaches the lower bound
(as could be said for $(n_0\lambda)^{-1}\ll t\ll v_0^{-1}$ and
$2<d<4$), only that it lies somewhere 
between the limits supplied by (\ref{limu}) and (\ref{limb}). In this
respect the case of $d=2$ and $t\gg v_0^{-1}$, $t\gg
(n_0\lambda)^{-1}$ is similar to that of $d=4$
and $(n_0\lambda)^{-1}\ll t\ll v_0^{-1}$. In addition both situations
retain the $t^{-1}$ mean
field decay rate, but with modified amplitudes, which for $d=2$ and
$t\gg v_0^{-1}$, $t\gg (n_0\lambda)^{-1}$ depend on $v_0$.

Note that in the limit $\tilde\lambda_{eff}^2\Delta/\bar v_0D^2\gg 1$,
the upper bound reduces to 
\begin{equation}
f\approx\sqrt{{D\Delta\sqrt{3}\over 16\pi v_0}}(Dt)^{-1},
\end{equation}
i.e. the upper and lower limits differ simply by a numerical factor of
$(2/\pi)^{1\over 2}$. On the other hand, if
$\tilde\lambda_{eff}^2\Delta/\bar v_0D^2\ll 1$, then $f\approx
(\tilde\lambda_{eff}t)^{-1}$.
In the limit of strong shear, where any reaction zones are broken up
almost immediately, we have recovered a mean field decay from this
upper bound. 

Finally, we can see that the lower bound (\ref{limb}) is of limited
usefulness in the small $\Delta$ or large $v_0$ limits, as the bound
decreases with either increasing $v_0$ or decreasing
$\Delta$. However, we can use the fact that 
$\tilde\lambda_{eff}^2\Delta/\bar v_0D^2$ is dimensionless in $d=2$
to obtain an improved expression for the densities, by performing a
perturbation expansion with this parameter. From (\ref{avf}) it
follows that the zeroth order
term of this series is a constant, equal to the small
$\tilde\lambda_{eff}^2\Delta/\bar v_0D^2$ limit
of the upper bound:
\begin{equation}
\langle a\rangle=
{1\over\tilde\lambda_{eff}t}\left[1+{\sqrt{3}\tilde\lambda_{eff}^2
\Delta\over 16\pi \bar v_0 D^2}+O\left(\left[
{\tilde\lambda_{eff}^2\Delta\over \bar v_0D^2}\right]^2\right)
\right].
\end{equation}

\section{Numerical Results}

In order to confirm some of our theoretical predictions we have
performed Monte-Carlo simulations in $d=2$.  Initially a square
lattice of size  
$L_x\times L_y$ was populated with equal numbers of randomly
distributed 
A and B particles. The evolution of this initial configuration was
simulated using the rare event dynamics (RED) technique (see also
\cite{me2}). 
In this Monte Carlo method, the time increment is determined by
the current configuration: if many changes in the configuration are
likely, then the time increment is small, whereas if the configuration
is very stable, 
the time increment is large. In RED, a list is made of all possible
changes 
to the configuration (events) together with the expected time
after which each event will occur (rates).  In the present model, two
distinct types of events could occur:

(1) A and B particles at $(x,y)$ could hop to neighbouring sites with
rates: 

\begin{tabular}{rl}
UP: & Rate$=1$ \\
DOWN: & Rate$=1$ \\
LEFT: & Rate$=1+s\left({y-0.5L_y\over L_y}\right)$ \\
RIGHT: & Rate$=1-s\left({y-0.5L_y\over L_y}\right)$, \\
\end{tabular}

\

\noindent where $v_0=2s/L_y$.
Note that the possible values of the shear gradient $v_0$ were
restricted to ensure that the hop rates remained everywhere positive
(i.e. $s\leq 2$).

(2) Each A particle could react with each B particle on the same
lattice site, with a reaction rate $\lambda$. The simulations also
allowed multiple occupation of each lattice site, in accordance with
our theoretical description of the system.

One step in a RED simulation consists of incrementing the time scale
with $\Delta t=1/\sum_j (r_j)$, where $r_j$ is the rate for event $j$,
and then allowing selection (and execution) of an event. The
probability 
that event $i$ is chosen is equal to $p_i=r_i/\sum_j (r_j)$.

For an efficient implementation the events are organized in a binary
tree,  
where each branch contains one event and has a weight equal to the
rate  
of that event. The weight of a parent node is equal to the sum of
the weights of its children. As the root node contains the sum of all
rates, the time increment $\Delta t$ is easily obtained: it is the
inverse 
of the weight of the root node. To select a particular event $i$ with
a probability proportional to its rate $r_i$, 
we start in the root node, descend to either its left or its right
child 
with a probability proportional to their weights, and iterate this
process 
until we have reached the bottom of the tree. The selected event is
then executed and
the weights in the tree of all events whose rates have changed, plus
their 
parents, grandparents, etc. are updated. The use of the binary tree
assures 
that the CPU time required for one step in the RED simulation scales
with 
the logarithm of the size of the tree.

In our simulations we have adopted periodic boundary conditions in the
direction of the shear flow (in the $x$ direction), but with hard wall
boundary conditions perpendicular to it (at $y=0$ and $y=L_y$).

The results of simulations on $800\times 200$ and $900\times 300$
lattices are shown in figures 8 and 9 respectively. Data for the
shearless cases ($s=0$) are averaged over $20$ runs, whereas those
with shear ($s=1$) are averaged over $40$ runs.
At very early times the decay is fairly slow (especially in the
$n_0=0.04$ case) as it takes time for the particles to ``find'' each
other. However at slightly later times we see a straight
line for the densities, which is quite well described by the
theoretically predicted $t^{-1/2}$ decay. Notice that the 
simulations both with and without shear are indistinguishable at these
early times, indicating that the shear really does have a
negligible impact early on (as was predicted in section 4). However
for $s\neq 0$, we
see a crossover at times of order $v_0^{-1}$ to a new regime,
which is fairly well described by a $t^{-1}$ decay.

Notice that at late times the densities in the shearless simulations
also tail off, and appear to fall away much more quickly than the
predicted $t^{-1/2}$ decay. Similar effects were also seen in the
original paper by Toussaint and Wilczek \cite{TW}, where the
$t^{-d/4}$ 
decay (for $d<4$) was first proposed. We interpret this result as
being 
an exponentially fast density decay caused by finite size effects,
which become appreciable when the segregated domain sizes become of
the same order as the system size. 
At large enough times these segregated zones will have grown
to a size where only two such regions exist (one A rich and the other
B rich). The depletion zones of these segregated areas will typically
be of the same order as the system length. Hence, if we have just two
zones remaining in a system of size $L\times L$, then the densities
would decay according to: 
\begin{equation}
{d\over dt}(L^2 a) \sim -JL \sim -(aD/L)L\sim -Da,
\end{equation}
where $J$ is the flux of particles flowing into the reaction
zone. This implies
\begin{equation}
a\sim \exp{(-\rm{(const.)}Dt/L^2)}.
\end{equation}
In order to test this prediction, we simulated an initially completely
segregated system of size $200\times 200$. The $t=0$ density profiles
for each of the two species were set up to rise linearly from zero
(around 
$y=L_y/2$) to a maximum on opposite boundaries (at $y=0$ and
$y=L_y$). This configuration mimics the effects of having a diffusion
length of the same order as the system size. The results of these
simulations (averaged over $5$ runs) are shown in figure 10, where an
exponential decay is found, in agreement with the above theory.

A further feature of the simulations concerns the effects of the hard
wall boundary conditions. Segregated zones neighbouring
the hard wall borders at the top and bottom of the system are
likely to be more stable, as for these zones fewer reactions are
occurring on their boundaries relative to similar regions in the bulk.
Hence, if the system is roughly square, then we might expect 
the reaction zone at late times (when only two segregated regions
remain) to lie roughly along $y=L_y/2$. This effect has indeed been
seen in our simulations in the case of zero shear. 

Finally, in figures 11--13, we show the results of various
``snapshots'' of 
an $800\times 200$ system as it evolves in the presence of a shear
flow. The effects of the shear in tearing apart the reaction zones are
clearly illustrated. 

\section{Conclusion}

In this paper we have studied theoretically and numerically the
effects of a linear shear flow on 
the two species reaction $A+B\rightarrow\emptyset$.
For $t\ll v_0^{-1}$ we found that the system was essentially
unaffected by the 
velocity flow, and the species again segregated into A and B rich
zones 
for $d<4$. In the regime $(n_0\lambda)^{-1}\ll t\ll v_0^{-1}$ and
with $Dt\gg n_0^{-2/d}$, our
results for the densities (to leading
order in a small density expansion) reproduced
those of the shearless case \cite{LC}. However at large times $t\gg
v_0^{-1}$, the critical dimension for the onset of
anomalous kinetics is reduced from $d_c=4$ to
$d_c=2$. 
Consequently, for $d\geq 2$, a mean field $t^{-1}$ decay
is adopted, though with a modified 
$v_0$ dependent amplitude. Notice, however, that
our calculations are for the case of equal diffusion constants. 
Although it would be possible to treat the case of $D_A\not
=D_B$, we believe (as was found in \cite{LC}) that such a modification
would be qualitatively unimportant, provided both diffusion constants
remain non-zero.

To go beyond the calculations presented in this paper, one could
attempt to generalise our analysis to the case of a 
non-linear shear flow. Unfortunately, excepting the very simple
arguments of section 2, it is not clear how to make analytic
progress in this case, as the equation for the Green functions
(\ref{Greq}) becomes very much harder to solve. 
However, it should prove possible to incorporate into our model a 
repulsive force between like particles. This would make our system
more similar to those considered in \cite{J,J1,IKR}, where a same
species hard core exclusion rule was imposed. In this way we could use
the field theoretic formalism to investigate the effects of exclusion
on reaction-diffusion systems with shear or drift.

\

\noindent{\bf Acknowledgments.}
\noindent  The authors would like to thank John Cardy for useful
discussions. MJH acknowledges financial support from the EPSRC, and
also 
{}from Somerville College, Oxford. GTB acknowledges financial support
{}from the EPSRC under grant GR/J78044, from the DOE under grant
DE-FG02-90ER40542, and from the Monell Foundation.

\newpage
\begin{figure}
\begin{center}
\leavevmode
\vbox{
\epsfxsize=5in
\epsffile{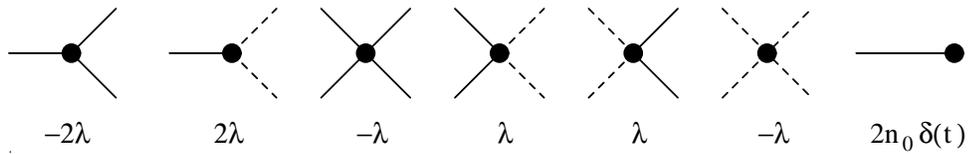}}
\end{center}
\caption{Vertices given by the field theoretic action.}
\end{figure}
\begin{figure}
\begin{center}
\leavevmode
\vbox{
\epsfxsize=5in
\epsffile{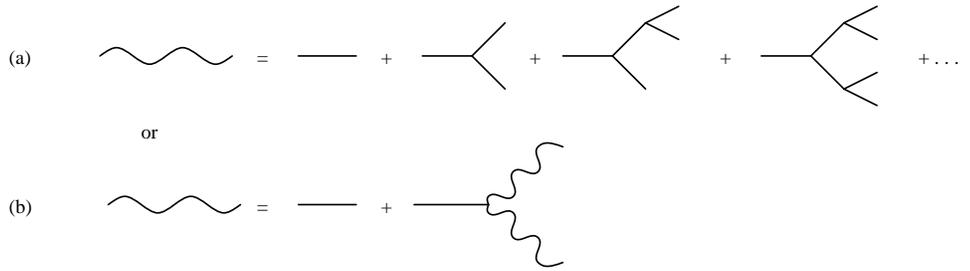}}
\end{center}
\caption{The classical density $\langle\phi\rangle_{cl}$, given by (a)
the sum over tree diagrams, or (b) an integral equation.}
\end{figure}
\begin{figure}
\begin{center}
\leavevmode
\vbox{
\epsfxsize=5in
\epsffile{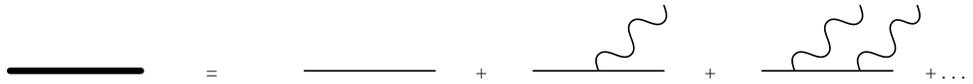}}
\end{center}
\caption{The diagrammatic sum for the $\phi$ response function.}
\end{figure}
\begin{figure}
\begin{center}
\leavevmode
\vbox{
\epsfxsize=5in
\epsffile{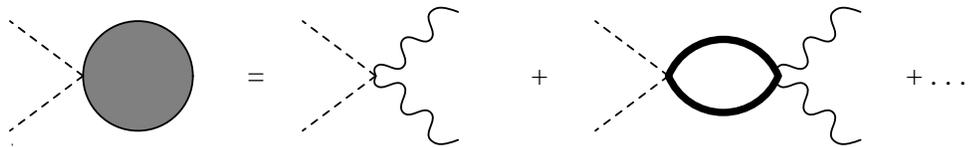}}
\end{center}
\caption{The initial term $\Delta$ is generated by the sum of diagrams
shown here. Only the leading order diagram is evaluated.}
\end{figure}
\begin{figure}
\begin{center}
\leavevmode
\vbox{
\epsfxsize=5in
\epsffile{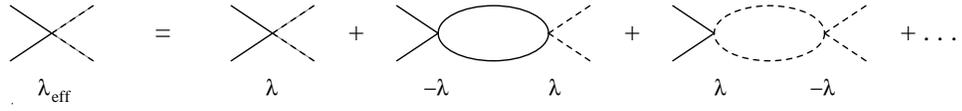}}
\end{center}
\caption{Diagrammatic expansion for the effective coupling constant.}
\end{figure}
\begin{figure}
\begin{center}
\leavevmode
\vbox{
\epsfxsize=5in
\epsffile{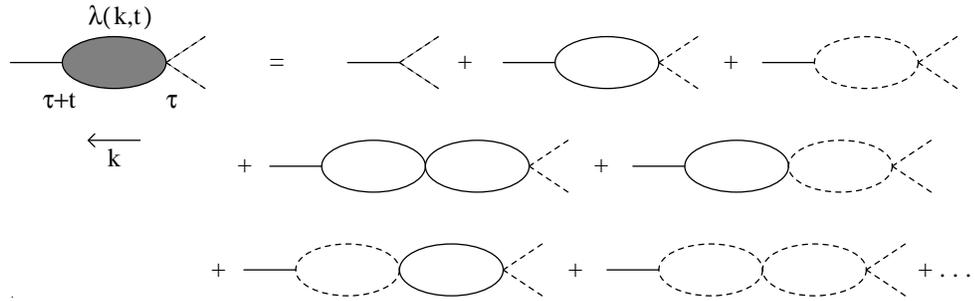}}
\end{center}
\caption{The sum of diagrams contributing to the primitively
divergent vertex function $\bar\lambda(k,t)$.}
\end{figure}
\begin{figure}
\begin{center}
\leavevmode
\vbox{
\epsfxsize=5in
\epsffile{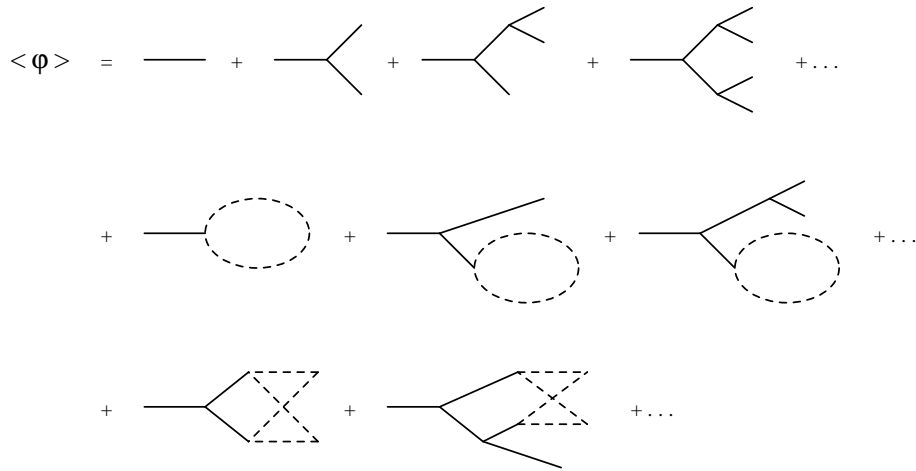}}
\end{center}
\caption{Diagrammatic expansion for $\langle\phi\rangle$, using the
initial terms $n_0$ and $\Delta$.}
\end{figure}
\begin{figure}
\begin{center}
\leavevmode
\vbox{
\epsfxsize=5in
\epsffile{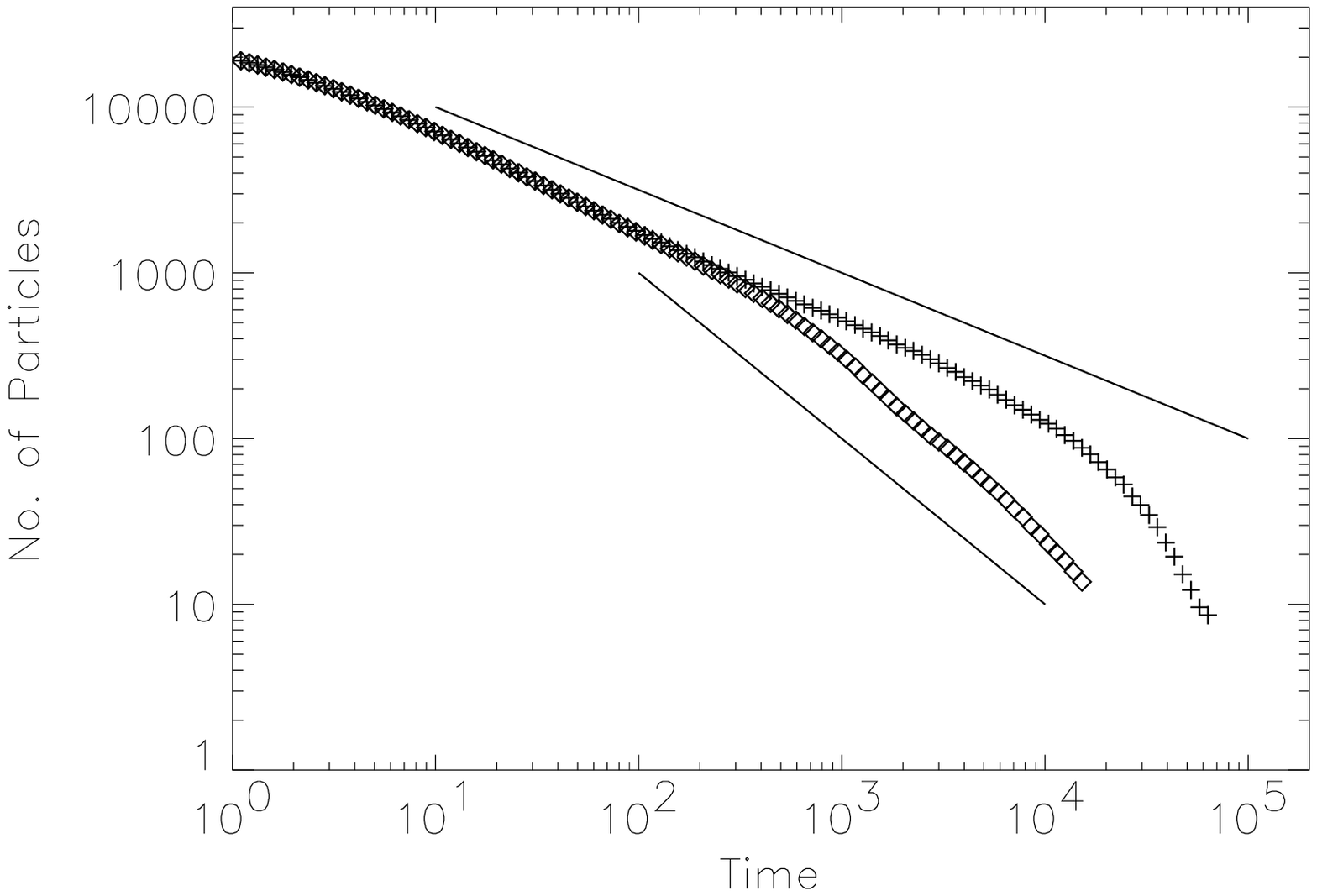}}
\end{center}
\caption{Simulation results on an $800\times 200$ system, with
$\lambda=1000$, $n_0=0.1$, and $s=1$ (diamonds) or $s=0$
(plus signs). The straight lines have gradients of $-0.5$ and $-1$
respectively.}
\end{figure}
\begin{figure}
\begin{center}
\leavevmode
\vbox{
\epsfxsize=5in
\epsffile{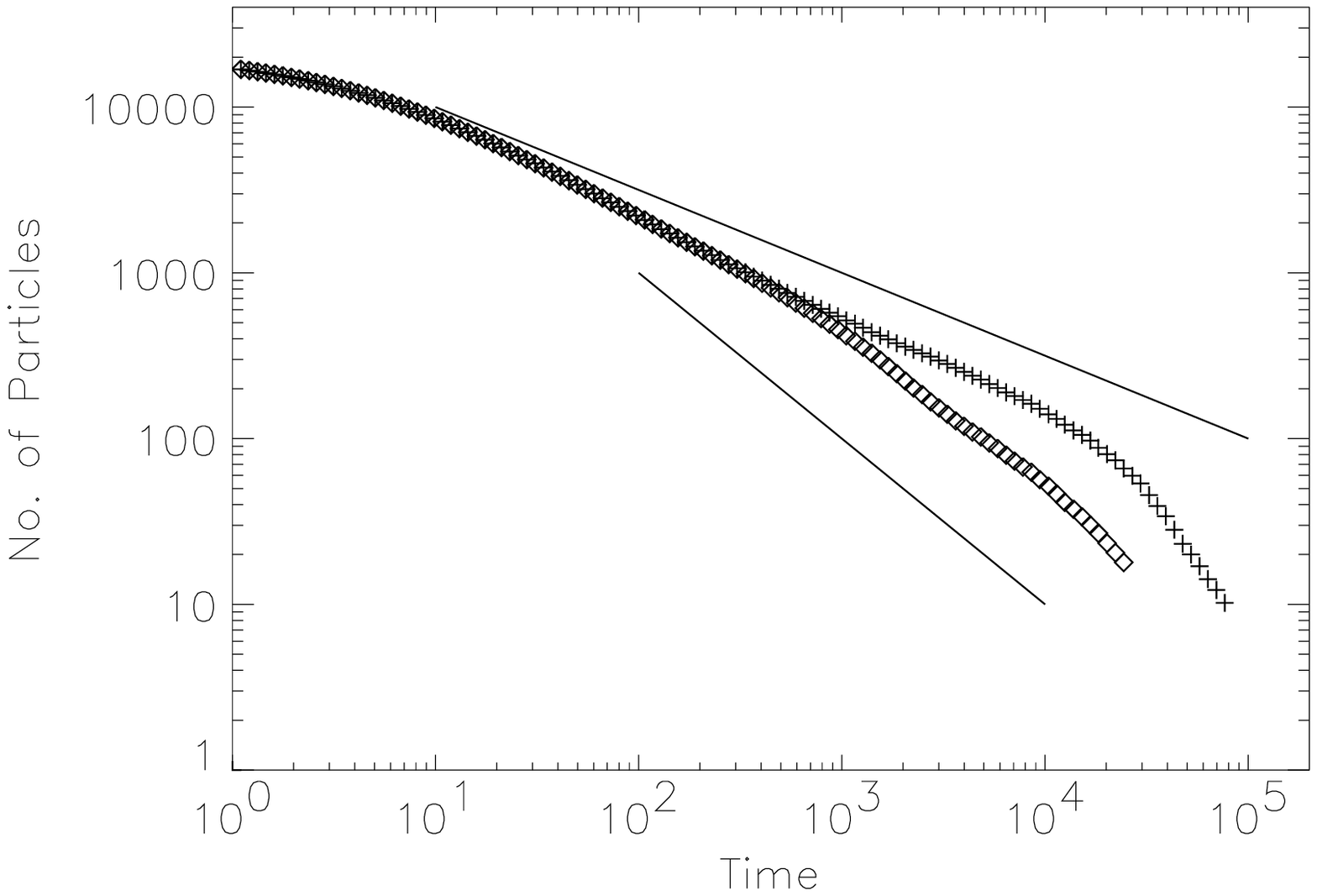}}
\end{center}
\caption{Simulation results on a $900\times 300$ system, with
$\lambda=1000$, $n_0=0.04$, and $s=1$ (diamonds) or $s=0$
(plus signs). The straight lines have gradients of $-0.5$ and $-1$
respectively.}
\end{figure}
\begin{figure}
\begin{center}
\leavevmode
\vbox{
\epsfxsize=5in
\epsffile{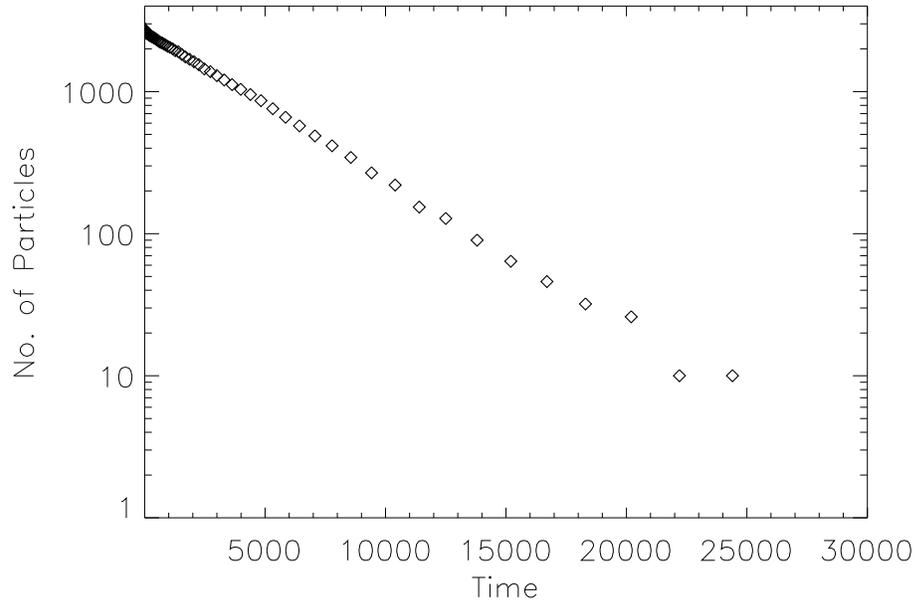}}
\end{center}
\caption{Simulation results on a $200\times 200$ system with
complete initial segregation and $\lambda=1000$, $s=0$.}
\vspace{3in} 
\end{figure}
\begin{figure}
\begin{center}
\leavevmode
\vbox{
\epsfxsize=3in
\epsffile{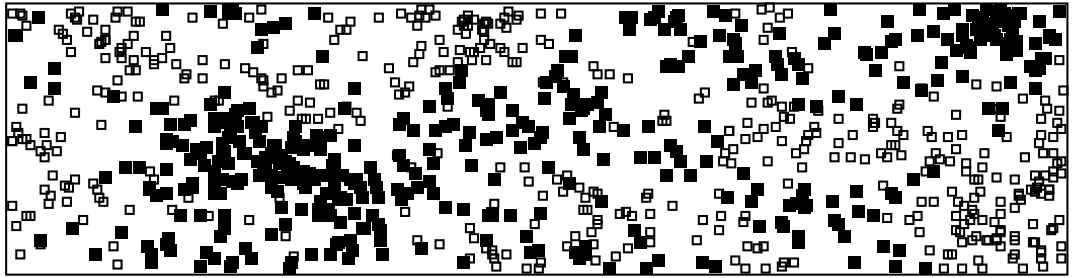}}
\end{center}
\vspace{-5.75in}
\caption{Snapshot of an $800\times 200$ system, with $s=1$, $n_0=0.1$,
$\lambda=1000$ at $t=300$.} 
\end{figure}
\begin{figure}
\begin{center}
\leavevmode
\vbox{
\epsfxsize=3in
\epsffile{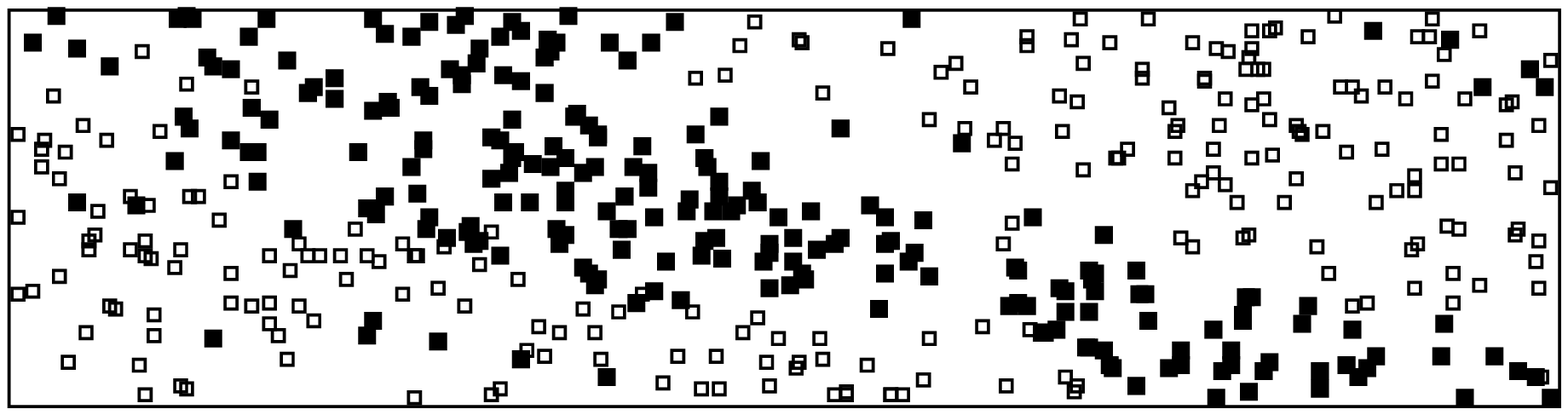}}
\end{center}
\vspace{-5.75in}
\caption{Snapshot of the same $800\times 200$ system at $t=900$.} 
\end{figure}
\begin{figure}
\begin{center}
\leavevmode
\vbox{
\epsfxsize=3in
\epsffile{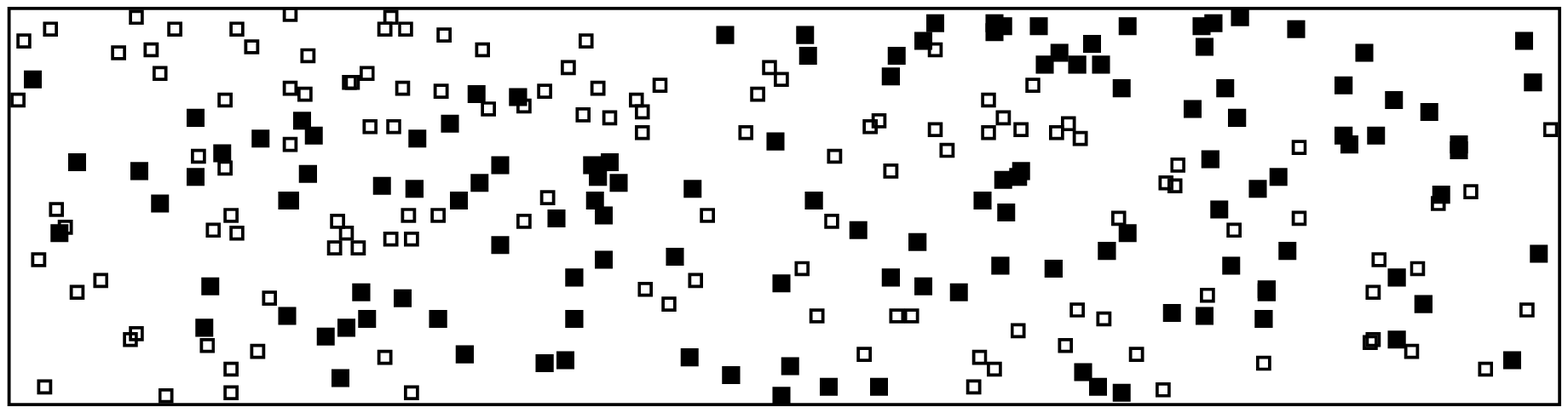}}
\end{center}
\vspace{-5.75in}
\caption{Snapshot of the same $800\times 200$ system at $t=1500$.} 
\end{figure}
\end{document}